\documentclass[12pt]{article}
\usepackage{amsfonts}
\usepackage{gensymb}
\usepackage{fullpage}
\usepackage{graphicx}
\usepackage{graphics}
\usepackage{mdwlist}
\usepackage{mathtools}
\usepackage{bbm}
\usepackage{hyperref}
\usepackage{algorithm2e}
\usepackage{amsmath}

\usepackage[natbibapa,nodoi]{apacite}
\newcommand{\Mod}[1]{\ (\mathrm{mod}\ #1)}

\usepackage{caption}
\usepackage{subfig}
\usepackage{graphicx}
\usepackage{fancyhdr}

\fancypagestyle{FirstPage}{
\lfoot{Copyright \copyright 2021 Shuqi Dai, Xichu Ma, Ye Wang, Roger B. Dannenberg}
}
\begin{document}
\date{May 8, 2021}
\newcommand{\beq}{\begin{equation}}
\newcommand{\eeq}{\end{equation}}
\newcommand{\bit}{\begin{itemize*}}
\newcommand{\eit}{\end{itemize*}}
\newcommand{\goal}[1]{ {\noindent {$\Rightarrow$} \em {#1} } }
\newcommand{\hide}[1]{}
\newcommand{\comment}[1]{ {\footnotesize {#1} } }
\newtheorem{lemma}{Lemma}
\newtheorem{theorem}{Theorem}
\newtheorem{proof}{Proof}
\newtheorem{defn}{Definition}
\newtheorem{algo}{Algorithm}
\newtheorem{observation}{Observation}

\title{Personalized Popular Music Generation Using Imitation and Structure}

\author{
  {\em Shuqi Dai}\\
  Carnegie Mellon University\\
  \texttt{shuqid@cs.cmu.edu} \\
  \and
  {\em Xichu Ma}\\
  National University of Singapore\\
  \texttt{ma\_xichu@u.nus.edu}\\
  \and
  {\em Ye Wang}\\
  National University of Singapore\\
  \texttt{wangye@comp.nus.edu.sg}\\
  \and
  {\em Roger B. Dannenberg}\\
  Carnegie Mellon University\\
  \texttt{rbd@cs.cmu.edu}\\
}

\maketitle

\begin{abstract}
    Many practices have been presented in music generation recently. While stylistic music generation using deep learning techniques has became the main stream, these models still struggle to generate music with high musicality, different levels of music structure, and controllability. In addition, more application scenarios such as music therapy require imitating more specific musical styles from a few given music examples, rather than capturing the overall genre style of a large data corpus. To address requirements that challenge current deep learning methods, we propose a statistical machine learning model that is able to capture and imitate the structure, melody, chord, and bass style from a given example seed song. An evaluation using 10 pop songs shows that our new representations and methods are able to create high-quality stylistic music that is similar to a given input song. We also discuss potential uses of our approach in music evaluation and music therapy.   

\end{abstract}

\section{Introduction}
\pagestyle{fancy}
\thispagestyle{FirstPage}
\lhead{\ifthenelse{\value{page}=1}{}{Dai, Ma, Wang and Dannenberg}}
\rhead{\ifthenelse{\value{page}=1}{}{Popular Music Generation (arXiv preprint)}}
\cfoot{\ifthenelse{\value{page}=1}{}{ \thepage }}
    \label{sec:intro}
    
Music is a universal language that has many formalizations, but there is no one guiding theory for music composition. Moreover, each person has unique music preferences. Our goal is to create personalized music automatically by imitating one or more songs that are selected by a user. To truly create new works, it is important to go beyond simple variation. Instead, we must capture, model and simulate a musical style. Given a style model, we can generate new music with enough novelty to be enjoyable, yet still retain similarity to the original.

One motivation for our work toward personalized music is the creation of music for Music Therapy. In particular, Rhythmic Auditory Stimulation (RAS) \citep{thaut1996rhythmic} is a useful neurological music therapy that helps people with Parkinson's disease to improve their gait performance and mobility. However, music pieces that meet both the criteria for RAS treatment and a patient's music preferences are hard to find. Limited therapist time impedes the selection and use of RAS in spite of expected benefits. 

An intuitive approach to personalized music is to imitate the music styles of a person's favorite pieces. 
Functional Magnetic Resonance Imaging (fMRI) data has revealed that familiarity is a crucial factor for emotional music engagement \citep{pereira2011music}.
A study of music choice also showed that familiarity with music positively predicts preference for songs, play lists, and radio stations \citep{ward2014same}.
We hope to contribute to more enjoyable and effective RAS for Parkinson's disease patients by capturing and automatically imitating patients' favorite songs. In addition, automatic generation of music for music therapy can avoid the copyright issues and access to large corpora of popular music, especially in communities with limited resources and budgets.

Recent practice in stylistic music generation mainly focuses on machine learning of general musical rules or style, which tends toward generic musical output with no support for personalization. The number of favorite and familiar songs that a person can provide is insufficient for deep learning approaches, which need large amount of training data. However, every piece of music has its own distinctive abstract qualities, for example, structure, melodic contour, rhythmic pattern, chord progression, bass line pattern, etc. These abstract qualities might vary a lot from piece to piece, even within a music genre and in works by the same composer. By focusing on distinctive musical qualities within the constraints of general rules of music, we hope to create more enjoyable music.

\begin{figure}[hbt!]
    \centering                                              \includegraphics[width=0.55\textwidth]{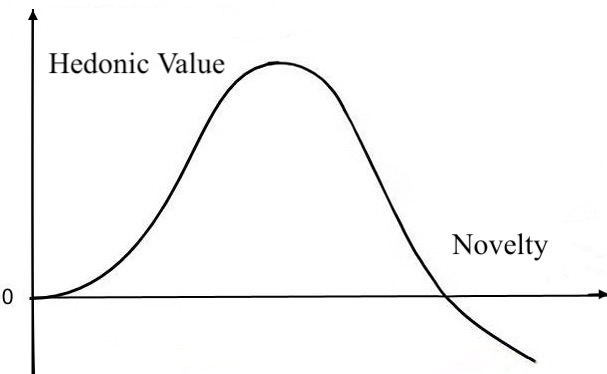}
    \caption{Inverted-U relationship: the Wundt Curve originally suggested by Wundt and later adapted by \cite{berlyne1971}, and the linking of favorability to familiarity/time curve by \cite{sluckin1983novelty}.}
    \label{iucurve}
\end{figure}

To avoid the problem of `smoothing over' distinctive attributes, and to further our goal of imitation, we need to `learn' from a single song that we call the \textit{seed song}. However, we cannot rely \textit{completely} on the seed song because that would result in too much similarity. Often, an overly similar song sounds as if the original were being played with mistakes or else the original was simply plagiarized. This is perhaps related to the `uncanny valley' phenomenon \citep{mori1970uncanny}. 
In addition, the inverted-U model of preference for music \citep{berlyne1971, sluckin1983novelty, chmiel2017back} suggests that the pleasantness (hedonic value) of a music piece increases as its novelty increases (familiarity diminishes), and will decrease once the novelty reaches a certain level (Figure \ref{iucurve}). Therefore, it is important to control how similar the new song should be in order to achieve both stylistic similarity and creative differences. This is a new challenge for the field of automatic music generation.

 Another significant issue in music generation is music structure, particularly at the level of phrases and sections. Longer term structure was at the heart of many early works on music generation, but largely ignored in more recent generation systems based on neural networks and sequence learning. The resulting failure of many systems to exhibit interesting long-term behavior has now been widely recognized and is receiving renewed attention. In this paper, we consider longer-term and higher-level music structure formed through section repetitions. For example a common pattern in popular music songs is \texttt{ABABB}, where letters stand for sections and repeated letters represent an approximate repetition of a section. 
 In addition to repetition structure, there can be a common rhythm in \texttt{A} and \texttt{B}, a repetition in melody contour from one phrase to the next, etc. In this paper, we consider both higher-level repetition structure and lower-level repetition music pattern structure in our representation and generation process.

We introduce a stylistic music generation model that is able to capture melody, chord and bass style from a single pop song and imitate them with structure information in a new complete piece. Like `music structure,' \textit{music style} is a general term that can refer to almost any aspect of music. Definitions are further complicated by the multi-level, multi-modal character of music representation---music can be notated and read, performed, and listened to, and each of these modalities has aspects of style \citep{dannenberg1993music, dai2018music}. In this paper, \textit{music style} refers to information at the symbolic music notation level, including rhythm, pitch and dynamics. We note that imitation of performance, orchestration, and production (especially in pop music) are also important aspects of style and perceived similarity, but we leave these to future work.

Our work focuses on popular music for multiple reasons. The original motivation for our work is to construct music for rhythmic auditory stimulation (RAS) for Parkinsons's disease patients. Millions suffer from this disease, and we conjecture that automation could assist clinicians and encourage more widespread use of RAS in music therapy. The second reason for studying popular music is the practical consideration that most people have familiarity and knowledge of popular music styles. Therefore, it is easier to compare, discuss, and evaluate popular music than other music. It is also the case that popular music has many conventions that seem to simplify the music generation problem. This is not to say that truly great popular music is easy to make, but at least we can formalize approaches that generate \textit{serviceable} popular music. 

Our work offers three main contributions. First, we offer methods that generate likable music overall. If music is not likeable, the fact that it is a successful imitation that listeners find \textit{similar} to something likeable is a small consolation. Second, we indeed produce music that listeners recognize as \textit{similar} to our seed songs. Thus, we are able to learn enough from a single input song to form an imitation, even when seeds vary from Chinese Pop to Western Pop songs.
Finally, one must ask if this whole enterprise of imitation to create likeable songs is valid to begin with. What if we make likeable imitations, but they are no more likeable than anything else?  We will show through correlation that an increased preference for the seed song predicts an increased preference for our generated imitation. Thus, our original idea to create personalized music by imitation shows promise.

Overall, we can say our methods generate music that listeners find likeable, and we can use imitation without large training sets to enhance listener preferences for generated music. In addition, our methods can generate complete customized songs of any length and containing a logical, hierarchical music structure, as opposed to generating a few bars of music or long rambling sequences lacking in longer-term structure. We will also see how individual models for stylistic melody, chord and bass generation can be combined to create hybrid styles. e.g. creating a song with melody style from song A, chord style from song B, and bass style from song C. While we describe one particular system in detail, we believe there are many ideas that can inform future systems, serve as a baseline for comparison, and offer insights into music perception and cognition.

In the remainder of this paper, we first look at related or successful stylistic music generation models (Section \ref{sec:survey}), and then introduce the data preparation process (Section \ref{fig:data_representation}) and system design details (Section \ref{sec:proposed}). In Section \ref{sec:experiments}, we describe both objective and subjective experiments to evaluate the model output. We conclude in Section \ref{sec:conclusions} with our findings and proposals for future research. More generated demonstrations can be found at \url{https://www.shuqid.net/demo-smg}.

\section{Related Work}
    \label{sec:survey}
    Music theory, music grammars, automata and Markov models are widely used in automatic music generation \citep{conklin1995multiple, lo2006evolving, thornton2009hierarchical}. Recently, deep generative models have become popular in this field  \citep{briot2019deep}. For stylistic music generation, current models mainly use the following approaches to model music style: 

1) Define style- and theory-related rules as constraints for
music generation \citep{hiller1985tsukuba, cope1991computers, cope2000algorithmic, cope2005computer}. Rules make the generation fast and 
controllable; however, the results may lack creativity and 
therefore sound less interesting. Also, the quality of the model
largely depends on the style/rule design. Since rules are often
hand-coded, it is hard to extend or adapt rules to create new styles. 

2) Learn music style implicitly from a large amount of training
data using deep learning models, especially sequence learning
models \citep{liang2016bachbot, hadjeres2017deepbach, huang2019counterpoint}.
While the results 
from this approach are impressive, they often lack common popular
hierarchical music structure such as 4-or-8-bar phrases or
repetition of melodic and rhythmic patterns. Recently,
representation learning
has motivated the practice of music style representation and
disentanglement using Variational Auto-Encoders
(VAEs) \citep{47078, yang2019deep, kawaiattributes} and Generative Adversarial Networks (GANs) \citep{yu2017seqgan}. 
These models are more controllable than previous deep learning
approaches, e.g. VAEs can be used to interpolate between phrases or explore near neighbors along different learned dimensions. However, the latent space still remains unexplained, and these models can hardly generate complete music
with hierarchical structure information. Often, the results are of fixed length.

3) Encode long-term music structure information using 
the Transformer, a sequence model based on self-attention 
\citep{vaswani2017attention}. This approach has been applied to stylistic
music generation tasks \citep{huang2018music, musenet, hakimi2020bebopnet}.
However, this approach requires a huge amount of training data
and computation power, and so far has only been used for general
music models because it would be difficult to find enough data to model of a specific style. The Transformer architecture
does not offer explicit representations of abstract music 
qualities such as chords, bass lines, or rhythms, so there is
no straightforward way to control or bias the model to produce
certain scales, hierarchical structure, repetition or even to
limit repetition. This approach lacks explainability as well. 

\cite{elowsson2012algorithmic} created an algorithmic composition system for popular music using probabilistic methods guided by music theory. Users can specify input settings like phrase length, chord transition matrix, and metrical salience histogram. The algorithm outputs a piece of music in MIDI format with melody, chord progression and simple phrases. They create a structure indicating phrase durations, accents within phrases, and phrase repetitions.
Chords are then selected using a Markov chain. Melody generation is based on ten different functions that estimate conditional probability of the next pitch and rhythm based on the previous note, current chord and accent locations, which are selected during the design of the overall structure. 
Humans rated the musical output slightly lower than human compositions, but not significantly.
This system generates a generic popular music style, but cannot represent style variations, 
personalize music, or model bass style.
We adapted some of the ideas in this work, for example, using theory and statistics to help capture and apply abstract qualities of style.

\section{Data Representation and Pre-processing}
    \label{sec:data}
\begin{figure}[!htbp]
    \centering
    \includegraphics[width=0.95\textwidth]{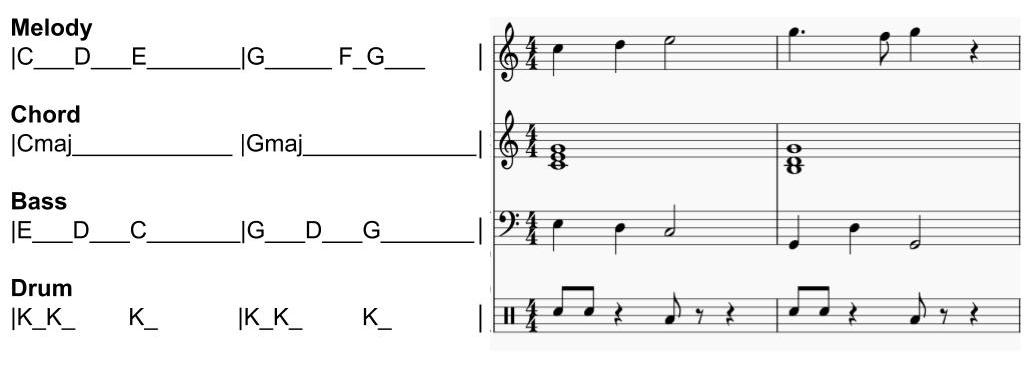}
    \caption{String-based data representation and its corresponding music notation.}
    \label{fig:data_representation}
\end{figure}

    The input/output data is represented in quantized MIDI format. 
We ignore everything but notes and meta-data such as tracks and 
instruments. To simplify our system, we transpose everything to the key of C major or A minor, eliminating all accidentals (sharps or flats). Some critical music information such as structure 
and chord progression cannot be explicitly expressed in MIDI.
Therefore, we convert MIDI to and from a data representation shown schematically at the left side of Figure \ref{fig:data_representation}.  We restrict music to a 
time signature of 4/4, and the sixteenth note is the
smallest unit of time, thus each bar contains an array of 16 
sixteenth notes. 
For computations, we represent pitch as 
an integer in [1...15] representing the scale degree over 2 octaves, starting on C. Therefore, differences in pitch, e.g. $p_1 - p_2$ are diatonic intervals; e.g. we do not distinguish major from minor intervals.

We represent one chord for each bar, using a limited vocabulary of seven triads: Cmaj, Dmin, Emin, 
Fmaj, Gmaj, Amin, Bdim (I, ii, iii, IV, V, vi, vii{\degree}).

Each duration is represented by an integer in the range [1...16] denoting the number of $16^{th}$ notes. Thus `notes' are pairs of pitch and duration. In our representation, there are only 240 ($15$ pitches$\times 16$ durations) different notes, which makes it practical in music generation to estimate probabilities for every possible note.

We have developed systems to analyze music MIDI files 
automatically by extracting melody, chords, bass and structure \citep{Jiang2019MelodyII, dai2020}, but for 
work reported here, we performed analyses by hand. 
We collected 13 MIDI songs and labeled them with an 
accurate analysis.

\section{Method}
    \label{sec:proposed}
     In this section we describe how we model and generate music. We begin with the overall model framework and then illustrate each module in the system individually. 

\subsection{Framework}
\begin{figure}[!htb]
    \centering
    \includegraphics[width=0.9\textwidth]{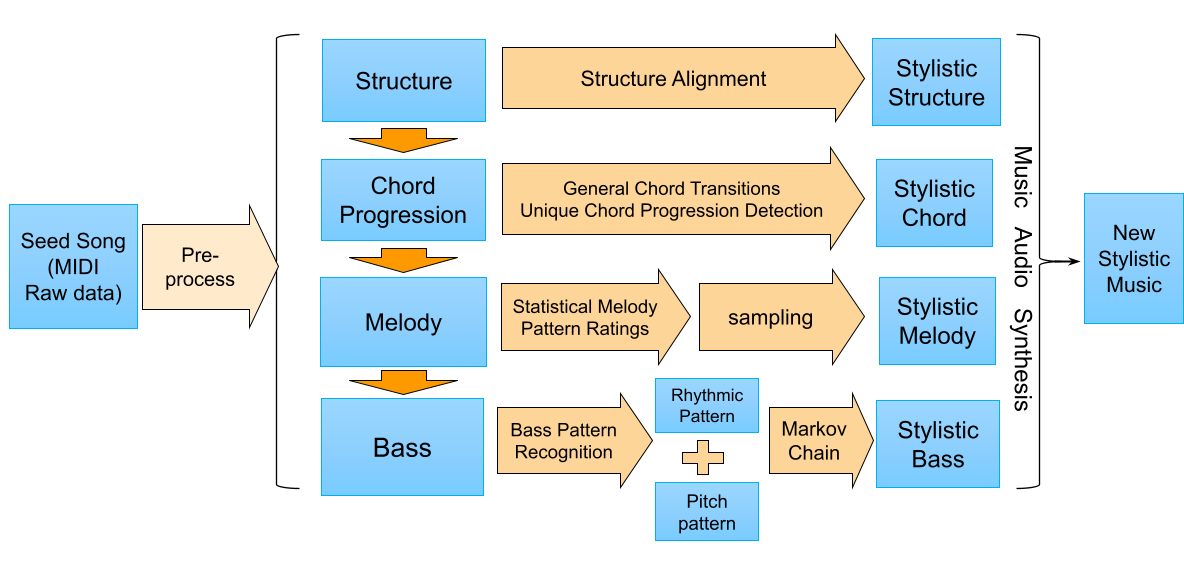}
    \caption{Framework of stylistic music generation system.}
    \label{fig:framework}
\end{figure}

We use statistical machine learning methods to imitate the styles of melody, chord progression, bass and structure from an input seed song. The system framework is shown in Figure~\ref{fig:framework}. The system takes one seed song as input. After some data pre-processing steps (described in section 3), we feed each part from the seed into a corresponding generation 
module.\footnote{Seed songs for different modules can be different, e.g. taking the melody from seed A and bass from seed B.} We also feed the results from structure and chord progression modules into the other modules as inputs.  The tempo is set to the same as the tempo of the seed song input. 
Finally, we combine the newly generated stylistic melody, chord accompaniment and bass MIDI tracks and synthesize them to obtain audio output. Detailed methods used for each module are described below.

\subsection{Structure Alignment and Generation}
We now discuss the higher-level structure of song sections. We describe structure using a list of tuples, where each tuple consists of a name (same sections or similar sections with slightly different variations share the same name), length (number of bars), and whether the section is a variation of the first section with the same name or not (`yes' means it is a variation, `no' means it is exactly the same section). For example, structure [[A, 8, no], [B, 10, no], [A, 8, no], [B, 10, no], [B, 10, yes]] means the song is divided into five sections `ABABB.' The first section is named `A' and has 8 bars; the last section is a variation of second section (named `B').   

\begin{figure}[!htb]
    \centering
    \includegraphics[width=0.7\textwidth]{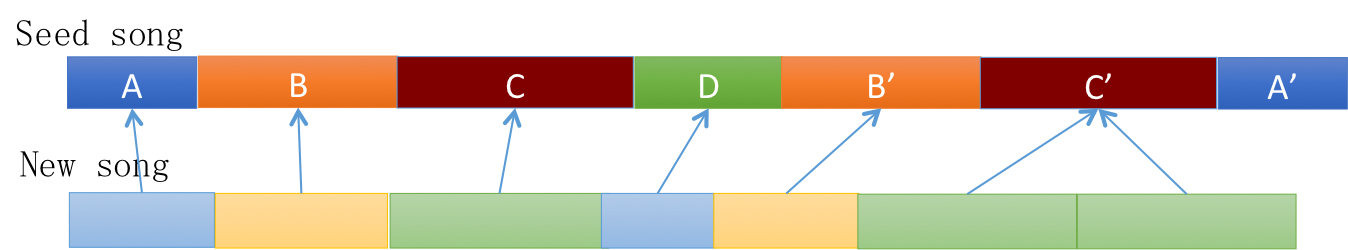}
    \caption{An example of ideal structure alignment.}
    \label{fig:structure}
\end{figure}

We developed three ways to generate new structures: (1) copy the seed song structure; (2) generate according to a specification string such as `AABABC' from the user and treat each section as 8 bars; (3) generate random structures, which includes selecting from a collection of typical structures. This gives flexibility in generation, e.g. for music therapy. Since the new structure can be different from the structure of the seed song, we need to align each new section to an appropriate seed song section. For example, in Figure~\ref{fig:structure}, we want to imitate the seed song shown at the top using the new structure shown below it. A good alignment will imitate the seed's intro style in the new intro, imitate seed's first chorus in the new chorus, etc. 

Since different songs have various structures and there is no standard solution for alignment, we deal with the problem in an unsupervised way using a nearest neighbor heuristic approach. We extract features $(L, P, M, F)$ from each section, representing length, number of sections before this one, whether it is a mirror/variation section or not, and the number of sections with the same section letter. We define the distance between two sections as follows, where the subscript 1 means a section from the seed song, 2 means a section from new song, $L_s$ and $L_n$ are the total bar lengths of seed song and new song, and $P_s$ and $P_n$ are the number of sections in the seed song and new song. $w_x$ are the weights of attributes, where $w_l = 0. 7 > w_p = 0.15 = w_m = 0.15 > w_f = 0.1$.

\begin{align}
    distance & = w_l|\frac{L_1}{L_s} - \frac{L_2}{L_n}| + w_p|\frac{P_1 + 1}{P_s} - \frac{P_2 + 1}{P_n}| + w_m|M_1 - M_2| + w_f|\frac{F_1}{P_s} - \frac{F_2}{P_n}|
\end{align}

We use nearest neighbor (NN) to find the best matching section in the seed for each new song section. However, if a new song section is not similar in those features to any of the seed song sections (we set a threshold for maximum section distance), we first try to align it with previous sections in the new song (also using NN). If
no previous section is similar (below threshold), then we consider this section as a completely new section with no seed reference.

\subsection{Stylistic Chord Generation}
To generate convincing chord progressions while imitating the harmonic style of the seed song, we combine statistical features from a general popular music data 
set\footnote{https://github.com/tmc323/Chord-Annotations}, seed song statistics, and distinctive sequences from the seed song. Within each section, we generate chord progressions in time order, selecting one or more successive chords at each step. Section endings are treated specially so that we can impose a stylistically consistent harmonic resolution to the section.

\paragraph{Blending Transition Matrix} Ignoring distinctive sequences for the moment, we create first-order Markov chain chord transition matrices from both general statistics and the seed song, then blend them together using $P_{trans} = (1 - \alpha) P_{general}+ \alpha P_{seed}$, where $\alpha$ is the tuning parameter. The combined transition probabilities are used to select the next chord up to the section ending sequence.

\begin{figure}[!htb]
    \centering
    \includegraphics[width=1\textwidth]{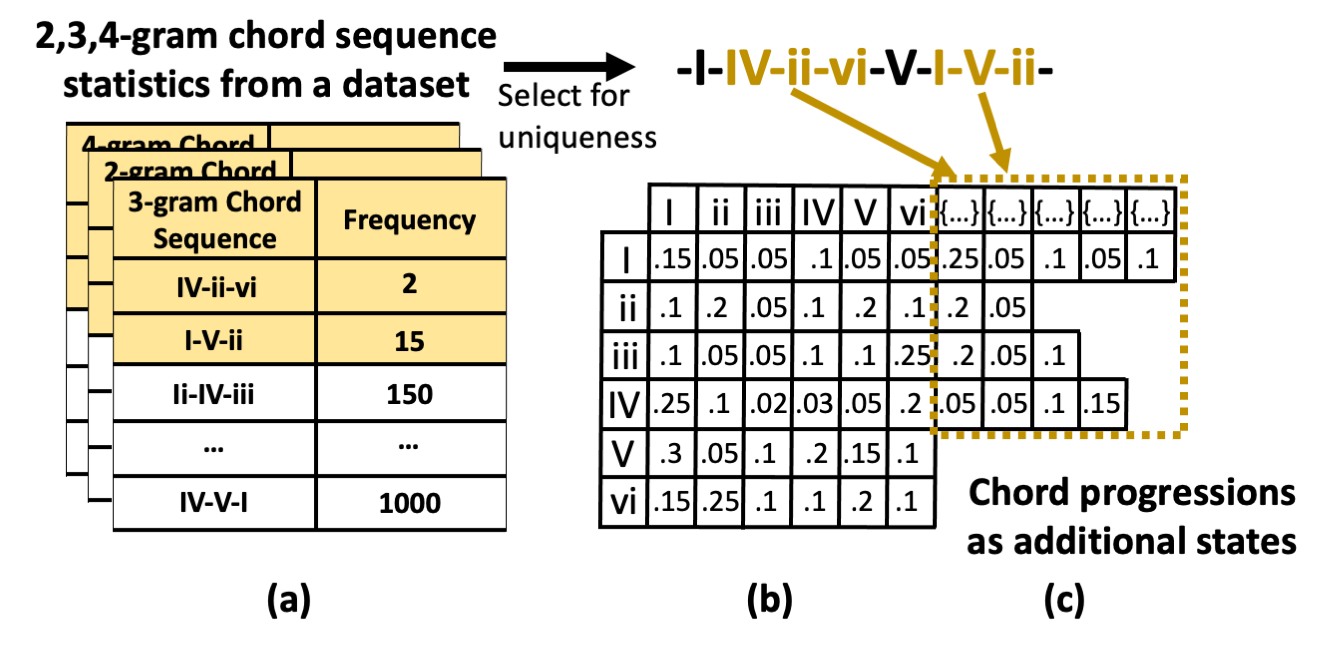}
    \caption{Distinctive chord sequence detection.}
    \label{fig:chord}
\end{figure}

\paragraph{Distinctive Chord Sequences in the Seed Song} Distinctive chord sequences are an important aspect of harmonic style and give songs a recognizable sound. We identify distinctive chord sequences in the seed song by comparing statistical features between the seed song and a general dataset consisting of 300 annotated pop songs. As shown in Figure \ref{fig:chord}(a), we find all of the 2,3,4-gram chord sequences in the seed song that occur less than a threshold ($5\%$) in the general statistics. We then extend our Markov chain approach (Figure \ref{fig:chord}(b)) by adding these chord sequences as possible new states (Figure \ref{fig:chord}(c)) resulting in an irregular-shaped transition matrix where some chords can transition to a short sequence. When such a transition occurs, the entire short sequence is output and generation continues from the last chord of that sequence.

\paragraph{Stylistic Cadence} Cadences are commonly used in popular music to end a phrase or section. Much like the handling of distinctive chord sequences, we calculate a cadence transition matrix that transitions from a chord to all cadence candidates (varying from a length of 2 chords to 6 chords). Beginning 6 chords from the end of a phrase, the normal transition probabilities are altered to include the possibility of finishing with a cadence sequence. Again, the cadence transition matrix combines both general statistics and seed song statistics, i.e. specific cadences in the seed song.

\subsection{Stylistic Melody Generation}
From music theory, we know that rhythm and pitch interact to form melody. Thus, we designed a number of melody style rating functions to evaluate the suitability of the next note given previous notes in the context of a chord progression. We generate melody in time order, note-by-note, using rating functions to estimate the probability of each possible next note and making a weighted random choice. Thus, we use a statistical sampling method. To avoid outliers, we generate 30 candidate melody sequences and pick the one with best rating as the stylistic melody output\footnote{In our experience, 
statistical sampling often leads to very unlikely selections
that seem to be musical mistakes, while making the most
likely choice at every step can become highly predictable
and repetitive. Generating many pieces and picking the 
most likely one rejects most songs with bad choices while
preserving variety and an interesting degree of surprise.}.

More formally, we have $n$ rating functions $P_i(x, y)$ where $x$ is the next note (pitch and duration pair), 
and $y$ is the context including the note position in the current bar and phrase, the previous notes, 
the chord progression, and the seed song. We treat these 
functions as independent probabilities, multiplying them together to form a weight for each note:
\begin{align}
w_x = \prod_{i=1}^{n} P_i(x, y)
\end{align}
Finally we select the next note at random based on the weights $w_x$. Note that it is not necessary for the weights produced by the $P_i$ to sum to one, and they are not all true probability functions. In the definitions that follow, we omit the usual normalization to produce proper probability functions since it is unnecessary.

The stylistic rating functions used to extract and represent melody style feature are inspired by  \cite{elowsson2012algorithmic}, who use similar methods to generate popular music. We extended these functions and use seed song statistics to encode specific melody style qualities from the input seed songs, and we developed additional rating functions that emphasize long-term structure. The rating functions consist of four categories: (1) rhythm-only ratings; (2) pitch-only ratings; (3) pitch and rhythm ratings; (4) long-term ratings. Similar to chord transition matrix blending in the previous discussion, almost all of the rating functions blend general statistics with seed song statistics, and we use $\alpha$ to represent the blending tuning parameter.

\subsubsection{Pitch style rating functions}
In the following, the notation $\lvert set\ description \rvert$ means the cardinality of the described set.
In general, we give details for probabilities based on the seed song, and we leave it to the reader to infer the formulas for \textit{general} probabilities.

\paragraph{Pitch frequency} Frequency of pitch $p$ occurring in both the seed song statistics and general popular music statistics:
\begin{align}
P_{freq}(p) & = \alpha P_{freq, seed}(p) + (1 - \alpha)P_{freq, general}(p),  \notag \\
\text{where }  P_{freq, seed}(p) & = \frac{\lvert\text{notes with pitch $p$ in seed song}\rvert}{\lvert\text{notes in seed song}\rvert}
\end{align}

\paragraph{Pitch harmony with chord progression} Given chord $c$, probability of pitch $p$ occurring in both the seed song statistics and general popular music statistics:
\begin{align}
P_{har}(p | c) & = \alpha P_{har, seed}(p|c) + (1 - \alpha)P_{har, general}(p | c), \notag \\
\text{where }  P_{har, seed}(p|c) & = \frac{\lvert\text{notes with pitch $p$ and chord $c$ in seed song}\rvert}{\lvert\text{notes with chord $c$ in seed song}\rvert}
\label{pitchharmony}
\end{align}

\paragraph{Pitch interval frequency} Frequency of consecutive pitch interval $p - p_{prev}$ occurring in both the seed song statistics and general popular music statistics.
\begin{align}
P_{interval}(p - p_{prev}) = \; & P_{interval}(\Delta p) = 
\alpha P_{interval,seed}(\Delta_p) + 
(1 - \alpha)P_{interval,general}(\Delta_p),\notag \\ \text{where }
 & P_{interval, seed}(\Delta_p) = \frac{\lvert\text{interval $\Delta_p$ in seed song}\rvert}
                                {\lvert\text{intervals in seed song}\rvert}
\end{align}

\paragraph{Pitch interval with harmony} Given chords $c_1$ and $c_2$, probability of the pitch interval $p - p_{prev}$ occurring in both the seed song statistics and general popular music statistics. This function is hand-crafted to capture multiple heuristics: (1) larger intervals are more likely to move to pitches that are more consistent with the chord, (2) larger intervals are also more likely to begin and end on chord tones, (3) larger intervals tend to go up, while smaller intervals tend to go down, (4) smaller intervals are more likely overall. 
We use $P_{har}(p|c)$ (Eq. \ref{pitchharmony}), derived from song statistics, to determine pitches that are consistent with chords, but we use rules to express interaction with intervals.

\paragraph{Downbeat pitch harmonic}
If a note starts on beat 1 or 3, then we
add more dependency on the harmony. Given chord $c$ for note with pitch $p$ and duration $d$ starting on beat 1 or 3, the harmonic rating is 
\begin{align}
    P_{downbeat}(p | d, c) = \begin{cases}
1.0,  & p \in \text{chord notes of } c, \\
(1 - \alpha) * 1.4 / dur + \alpha * P_{non, seed}(p | d, c), & p \notin \text{chord notes of } c \end{cases} \notag \\
\text{where } P_{non, seed}(p | d, c) = \frac{\lvert\text{non-chord tones starting at downbeat with duration $d$ in seed song}\rvert}{\lvert\text{notes starting at downbeat with duration $d$ in seed song }\rvert}
\end{align}

\subsubsection{Rhythm style rating functions}
We now consider probability distributions regarding rhythm and note duration.

\paragraph{Note duration frequency} Frequency of note duration (length) $d$ appearing in both the seed song statistics and general popular music statistics, 
\begin{align}
P_{durfreq}(d) & = \alpha P_{durfreq, seed}(d)  + (1 - \alpha) P_{durfreq, general}(d | tempo), \notag \\
& \text{where } P_{durfreq, seed}(d) = \frac{\lvert\text{notes with duration $d$ in seed song}\rvert}{\lvert\text{notes in seed song}\rvert}
\end{align}

\paragraph{Note duration transition} Probability of a note duration given previous consecutive note duration $d$ in both the seed song statistics and general popular music statistics,
\begin{align}
P_{durtrans}(d_1 | d_2) = & \alpha P_{durtrans, seed}(d_1|d_2) + (1 - \alpha) P_{durtrans, general}(d_1 | d_2), \notag \\
\text{where } P_{durtrans, seed}(d_1|d_2) & = \frac{\lvert\text{note transitions with duration $d_2$ to $d_1$ in seed song}\rvert}{\lvert\text{note transitions in seed song}\rvert}
\end{align}

\paragraph{Rest note duration}
Rest notes are more common to see in the last bar of a phrase/section. For rest notes in the middle of a phrase, we use both rest note duration statistics from the seed song and music theory rules to evaluate the rating.
\begin{align}
P_{durrest}(d|p = rest) = \begin{cases} 1.0, & \text{section end} \\
\alpha  P_{durrest, seed}(d|p = rest) + (1 - \alpha) * 1 / d, & \text{otherwise}
\end{cases} \notag \\
\text{where } P_{durrest, seed}(d|p = rest) = \frac{\lvert\text{rest note with duration $d$ in seed song}\rvert}{\lvert\text{rest note in seed song}\rvert}
\end{align}
    
\paragraph{Note onset position and duration} Originally, we used a general rule that only prevents syncopation at the $16^{th}$ note level. For imitation, we want to capture more detail from the seed song, so we estimate the probability of a note duration given each note onset position in $16^{ths}$ within a bar ($onset_p \in [0...15]$). Since the seed song has relatively few notes, we use a form of smoothing. E.g. position 4 relative to half notes is equivalent (modulo 8) to position 12, relative to quarter notes (modulo 4) is equivalent to 0, 8 and 12, and relative to eighths (modulo 2) is equivalent to 0, 2, 6, 8, 10, 12 and 14. We use a weighted sum of the statistics for all these positions.

\begin{align}
    P_{pos}(d | onset_p) = & \alpha \Big[\sum_{i = 1}^{4} \frac{2^i}{16} \sum_{onset_p \equiv j \Mod{2^i}} P_{pos, seed}(d | j)\Big] + (1 - \alpha) P_{pos, general}(d | onset_p), \notag\\
    P_{pos, general}(d | onset_p) = & \begin{cases}
    0,  & onset_p \equiv 1 \Mod{2} \quad \& \quad d \equiv 0 \Mod{2} \\
    1, & otherwise
\end{cases}.
\end{align}

\subsubsection{Pitch and Rhythm dependent rating functions}

\paragraph{Harmonic and note duration}
Notes longer than quarters are more likely 
to be chord tones. Notes shorter than quarters are more likely to be non-chord tones. If we use $P_{har}$, based on song statistics, to capture the notion of chord tones, the following rather obscure equation captures this heuristic:
\begin{align}
P(p, d | c) = ((P_{har}(p|c) - 0.5) * \log_2(d / 4) + 1.0) / 2.0.
\end{align}

\paragraph{Pitch Interval and note duration} Larger intervals occur after longer note durations. We combine a probability estimated from a melody collection with seed song statistics. 
\begin{align}
P_{di}(p_2 - p_1 | d_1) & = P_{di}(\Delta p | d_1) = 
\alpha P_{di,seed}(\Delta p | d_1) + (1 - \alpha) P_{di,general}(\Delta p | d_1), \text{where } \notag \\
P_{di,seed}(\Delta p | d_1) & = \frac{|\text{intervals $\Delta p$ with first note duration $d_1$ in seed song}|}{|\text{intervals with first note duration $d_1$ in seed song}|}.
\end{align}

\paragraph{Notes spanning chord changes}
Usually, new notes begin at chord changes. We do not 
allow a note to span two chords unless this occurs 
in the seed song.

\paragraph{Last bar and the tonic}
Sections ending with a strong perfect authentic 
cadence almost always end on the tonic (i.e. the pitch C) 
in popular music. 
The ending note is more likely to have longer duration than shorter notes too. This rule-based constraint overrides probabilistic sampling in our music generation model.

\subsubsection{Long-term Melody Structure}

\paragraph{Melody Contour}
\begin{figure}[!htb]
    \centering
    \includegraphics[width=0.92\textwidth]{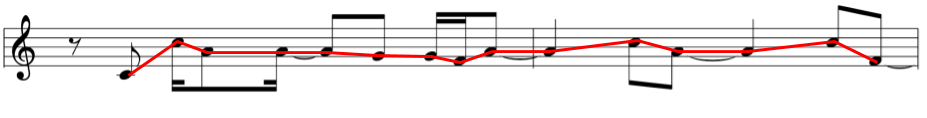}
    \caption{Red line represents the melody contour.}
    \label{fig:melodyc}
\end{figure}

As shown in Figure \ref{fig:melodyc}, melody contour (melody shape, configuration, outline) describes the general shape of the melody and proves to be very useful for identifying the melody style \citep{adams1976melodic}. In order to imitate the the melody contour in the seed song, we calculate a \textit{melody pitch contour similarity}, which is based on a Dynamic Time Warping (DTW) \citep{berndt1994using} algorithm. First, we transform two melody note sequences by taking the pitch or rest at each 16$^{th}$ note. E.g. a note with pitch $p$ and duration 4 in the original melody will become 4 consecutive copies of $p$ in the new sequence. We compute the edit distance using a substitution cost based on similar absolute pitch difference and similar melodic direction. We normalize the DTW distance and subtract from 1 to get a similarity.
\begin{align}
cost_{sub}(p_1, p_2) = \; & w_1 \cdot dist_{pit}(p_1, p_2) \notag\\
+ \; & w_2 \cdot dist_{dir}(prev_1 - 
p_1, prev_2 - p_2), \notag \\
\text{where } & \text{$w_1 = 1.0$ and $w_2 = 2.0$ are weights } \notag\\
\text{ and } & prev_i \text{is the pitch before } p_i.
\end{align}

\begin{align}
dist_{pit}(p_1, p_2) = \begin{cases}
12, & \text{$p_1$ or $p_2$ is a rest,}\\
|p_1 - p_2|, & otherwise.
\end{cases}
\end{align}

\begin{align}
dist_{dir}(\Delta_1, \Delta_2) = \begin{cases}
0, & \Delta_1 = \Delta_2 \text{ or } \Delta_1 \cdot \Delta_2 > 0 \\
2, &  \Delta_1 \cdot \Delta_2 < 0 \\
& \text{ or $\Delta_1$ or $\Delta_2$ is a rest}\\
1, & otherwise.
\end{cases}
\end{align}

The insertion cost for adding a pitch or rest $p_X$ in sequence $X$ compared to sequence $Y$ is defined as:
\begin{align}
cost_{ins}(p_X, Y) = & \begin{cases} |p_X - \overline{p_Y}|, & p_X \text{ is a pitch } \\ 10, & p_X \text{ is a rest,}\end{cases} \notag \\
\text{ where } & \overline{p_Y} \text{ is the average pitch of } Y.
\end{align}

The DTW calculation for the distance between sequences $X_i$ and $Y_i$ is
\begin{align}
dist(X,Y) = dtw( & |X|,|Y|), \text{ where } \notag \\
dtw(i, j) = min( & dtw(i - 1, j - 1) + cost_{sub}(X_i, Y_j), \notag \\
& dtw(i-1, j) + cost_{ins}(X_i, Y),  \notag \\
& dtw(i, j - 1) + cost_{ins}(Y_i, X)).
\end{align}

We are generating a new melody left-to-right, so matching a few notes to the entire seed melody contour might incur an extra cost proportional to the difference in lengths. We avoid this problem by matching only the prefix of the seed contour that gives the lowest DTW distance.

Finally, to get from the DTW distance to a similarity, we divide the DTW distance by the distance to a `flat' melody consisting only of the average seed melody pitch, giving a value from 0 (perfect match) to 1 or more (very poor match). We subtract this from 1 and limit the lowest value to zero to get a similarity:
\begin{align}
& P_{sim}(X, Y) = max(1 - \frac{a}{b}, 0)\text {, 
where } \notag\\
& a = dist(X, Y), b = \text{`poor match' distance.}
\end{align}

\paragraph{Rhythm Onsets Distribution Similarity} Apart from the rhythm style ratings, we also compare the rhythm onsets distribution of the generated imitation to the original seed. The \textit{rhythm similarity metric} is based on matching note onset times. Onsets times are quantized to 16$^{th}$ notes. Given two melodic sequences we define $similarity_{rhythm}$ to be the proportion of 16$^{th}$ note offsets where both sequences or neither sequence contains an onset (i.e. \textit{accuracy}).
With sets of onset times $A$ and $B$ (with times measured from the sequence beginnings), and sets of non-onset times $A^C$ and $B^C$, we define rhythm similarity as
\begin{align}
similarity_{rhythm} = \frac{|A \cap B| + |A^C \cap B^C|}{|A| + |B| + |A^C| + |B^C|}.
\end{align}
As each new duration is considered, we compute this similarity, and if the similarity is below a threshold, we disallow that duration choice.

\subsection{Stylistic Bass Generation}
We represent bass style using patterns. First, each bar of bass is represented by a rhythm pattern that denotes onset times. E.g. \texttt{[0\_\_\_,0\_\_\_,\_\_\_\_,0\_\_\_]} denotes onsets on beats 1, 2 and 4. For each pattern, we form a \textit{Chord Tone Frequency Matrix} with one row for each onset (in this case the rows represent beats 1, 2 and 4). The 4 columns are the probabilities of the bass note being the root, third, or fifth of the chord or a non-chord tone. We also form a \textit{Chord Tone Transition Matrix} for each onset where the columns are the probabilities of the \textit{next} bass note being the root, third, fifth or non-chord tone. We use the most frequent rhythm pattern in the seed song section together with the two chord tone matrices to generate bass.

\paragraph{Pattern-based bass generation}
We generate the stylistic bass section-by-section. 
The output bass rhythm copies the rhythms of the first and last bars of the seed song, and every other bar uses the most frequent bass rhythm pattern. Bass note selection follows the \textit{Chord Tone Frequency} and \textit{Chord Tone Transition} matrices. These are used to estimate the probability of $\mathbf{p}$, a vector representing each bass note as root, third, or fifth. We use the Viterbi algorithm to find the $\mathbf{p}$ with the maximum likelihood according to this model (equation \ref{bassgeneqn}).

This is a simple model of bass lines, but it does capture general stylistic characteristics such as always playing the root, alternating root and fifth, or playing arpeggios. It also uses the most common rhythmic pattern, which is a very important aspect of the bass style.

\begin{align} \label{bassgeneqn}
    \mathbf{p} & = argmax_{\mathbf{p}} \prod_i [P(p_i)P(p_i|p_{i - 1})], \quad p_i = \text{root, third, fifth, other}
\end{align}
\begin{figure}[!htb]
    \centering
    \includegraphics[width=0.6\textwidth]{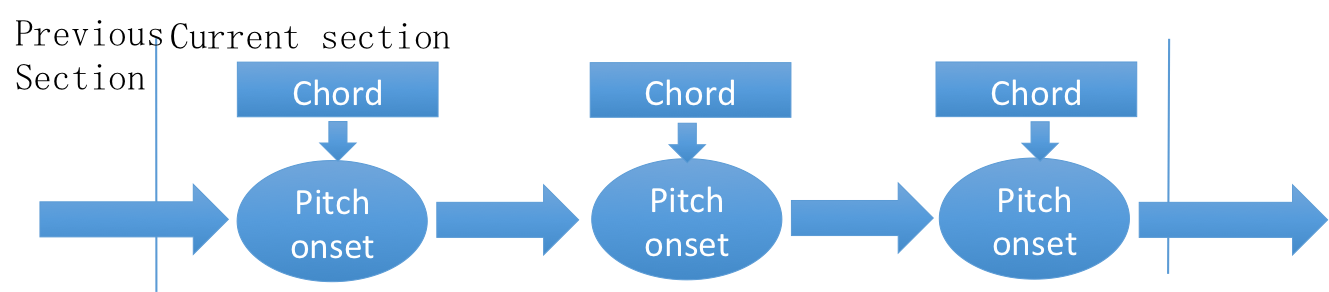}
    \caption{Pattern-based bass generation model using current chord for context and transition matrix for note-to-note probabilities.}
    \label{fig:bass}
\end{figure}

\section{Evaluation}
    \label{sec:experiments}
    We conducted both objective and subjective evaluations of our stylistic music generation system. We collected 13 MIDI songs and analyzed them by hand. Three of them are used in the training stage for parameter tuning. The other ten songs, five Western pop songs and five Chinese pop songs, are used as seed songs for evaluation. These test songs are quite popular and recognizable. One application of our system is to imitate Parkinson patients' favorite pop songs, so we wanted to imitate songs that are already popular and familiar. Our results show that our stylistic music generation system is able to create music with both high musicality and similarity to the seed song. We further explored the factor of familiarity in music generation.

\subsection{Objective Evaluation}

Objective evaluation of music is not well-understood. In fact, if we had a perfect evaluation function, music
generation would become a problem of search and optimization.

Similarity measures are also quite limited. An exact copy of a melody is the most similar one, but our goal is to create \textit{stylistically} similar music \textit{without} making a direct copy. Objectively, we measure success by showing that our many measures of similarity are satisfied. For melody and chord generation, we used a paired t-test to compare the estimated probability in terms of our multiple rating functions. The results show that the estimated probability of stylistic imitations is significantly higher (p-value $< 0.001$) than the unconstrained generation or imitations of other seed songs. In other words, our generated similar songs are more similar to their seed song than to other songs, and generated similar songs are more similar than unconstrained generated songs, in terms of our objective functions.

We do not want similarity to come at the cost of overall rating. We should at least expect the generated songs to have approximately the same probability as their seed songs in terms of our rating functions. A paired t-test shows that there is no significant difference in estimated probabilities (objective rating functions) between the original seed songs and the generated imitations (p-value $= 0.211$). Moreover, we find that $57\%$ of the generated imitations have higher probability than their seed songs. Thus, our approach can produce music that is similar to human-composed music in terms of our objective rating functions. This is consistent with subjective evaluations discussed in the next section. 

\subsection{Subjective Evaluation}

\subsubsection{Evaluation Design}
\begin{figure}[!htbp]
    \centering
    \includegraphics[width=1.0\textwidth]{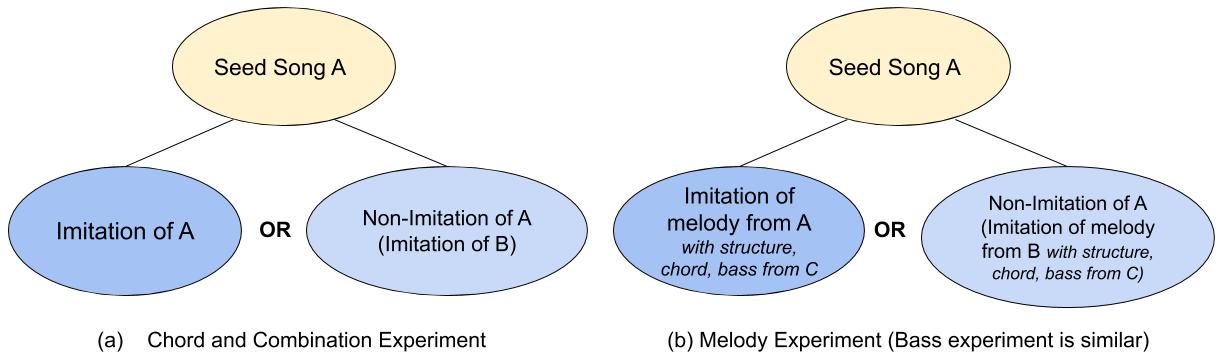}
    \caption{Settings for each pair of example in the experiment.}
    \label{fig:seed}
\end{figure}

We conducted human listening evaluations for the melody, chord and bass modules in our system and for the combination of all three. Listeners first answered
questions about their music background, for example, their
music practice experience, music genre preference, age and
gender. Then, the listeners answered four parts of the
evaluation (melody, chord, bass and combination) in a random order. In each part, listeners compared and rated ten example pairs.  Within each example pair, there are
two music fragments to listen to, each of them around 30
seconds in length. As shown in Figure \ref{fig:seed}, one fragment is the seed song. The other is either a stylistic imitation based on the seed song (called an \textit{imitation}) or an imitation of some
other seed song (called a \textit{non-imitation}). We asked users to 1) indicate if they have heard the music fragments before the listening survey; 2) judge the similarity between the two music fragments in each example pair; 3) rate how much do they like the music fragments. Both similarity and preference score ranges are the integers from 1 to 5. Every user will randomly get either an imitation or a non-imitation version for each seed song (seed song and the generated song are displayed in random order), and for each part, there are five imitations and five non-imitations. The ten seed songs are also presented in random order for each part. Users must listen to the full length of the songs, otherwise their ratings are not counted.

As shown in Figure \ref{fig:seed}(b), for the melody study, we want the similarity to the seed song to depend only on whether the melody imitates the melody of the seed song (A), or another song (B). To remove any coordination with chords, bass and structure, we derive these non-melodic elements by imitating a third song (C). To illustrate the problem this solves, imagine if the imitation uses the same chords as the seed song. The listener might rate the melody as similar merely because of a similar chord progression. In addition,
the combination of melodic contour and chords might be enough to substantially reproduce the seed song melody. We remove this information by taking chords from another song. 
For the bass similarity study we take a similar approach, 
imitating the melody, chords and structure from a third song (C) in both the \textit{imitation} and \textit{non-imitation} cases. In order to help non-musicians identify melody and bass, we give examples of what is a melody or bass line before the evaluation. For the chord similarity study (Figure \ref{fig:seed}(a)), we present only simple block chords from the songs with no melody or bass. 
For the combination study, we imitate
all parts including melody, chord, and bass. 

We use MIDI and software synthesizers to produce audio
versions of the songs. Within any pair of song fragments,
the synthesizer settings are identical so that differences
in sound quality are not a factor in similarity ratings.
We also adjust song settings such as pitch range, key signature and tempo within each example pair in order to avoid their influence. Here we use the original production and song settings of the seed song for each pair. 

\subsubsection{Results and Discussion}
\begin{figure}[!t]
\centering
\subfloat[Gender]{
  \includegraphics[width=0.2\linewidth]{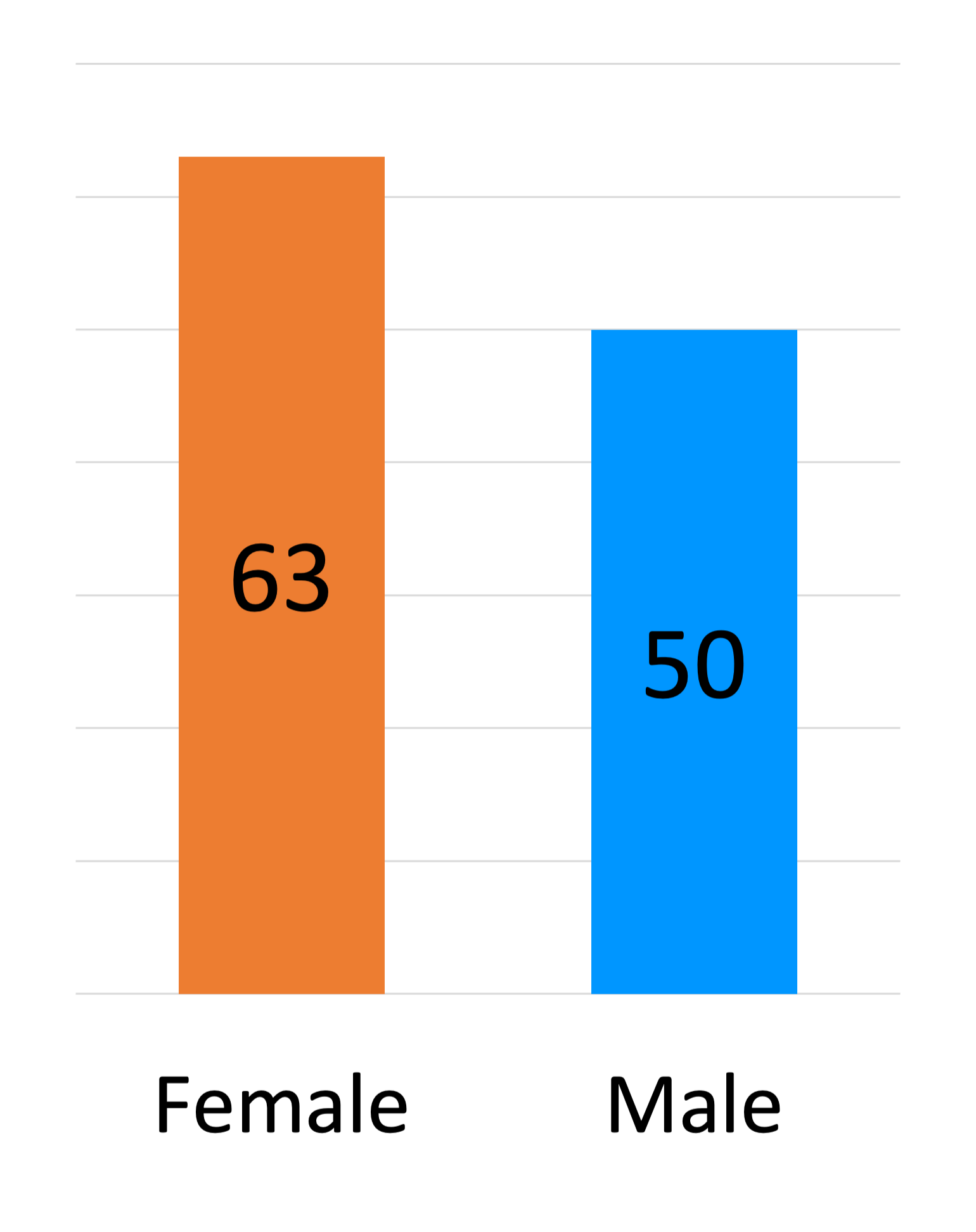}
}
\subfloat[Age. X-axis are different age groups ]{
  \includegraphics[width=0.4\linewidth]{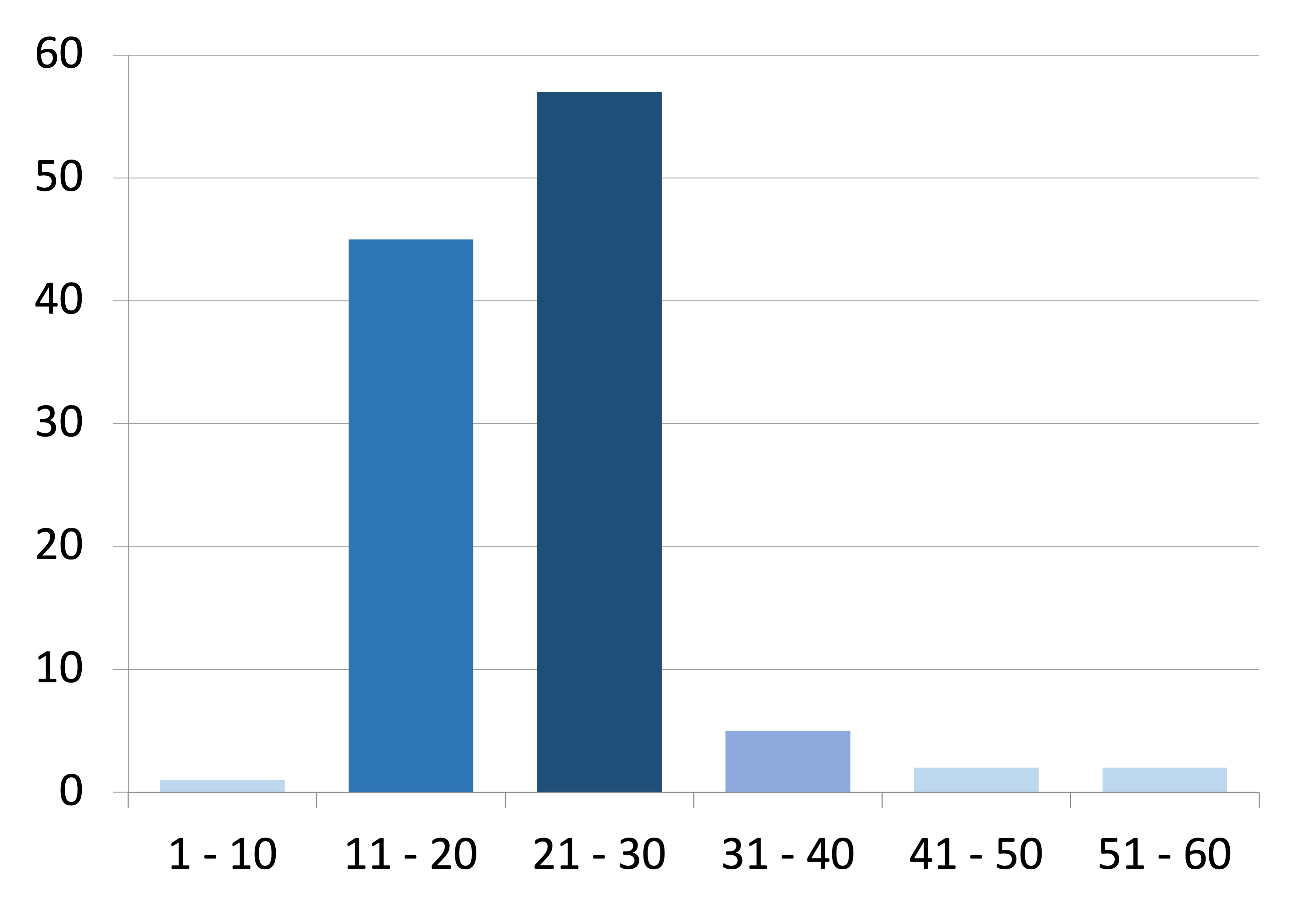}
}
\subfloat[Music Level. L5 means expert]{
  \includegraphics[width=0.35\linewidth]{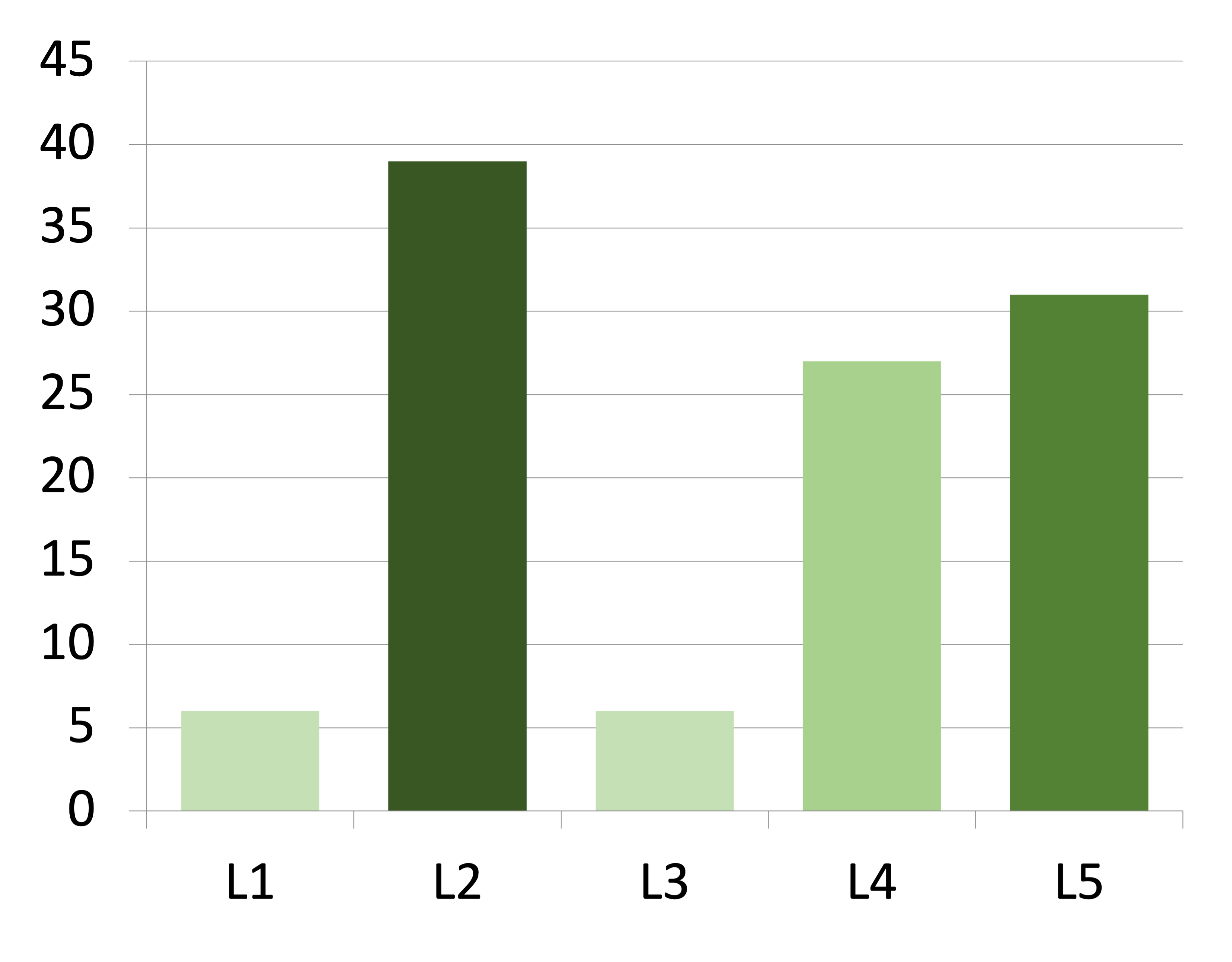}
}
\newline
\subfloat[Music Preference. Users can select multiple genres]{
  \includegraphics[width=0.5\linewidth]{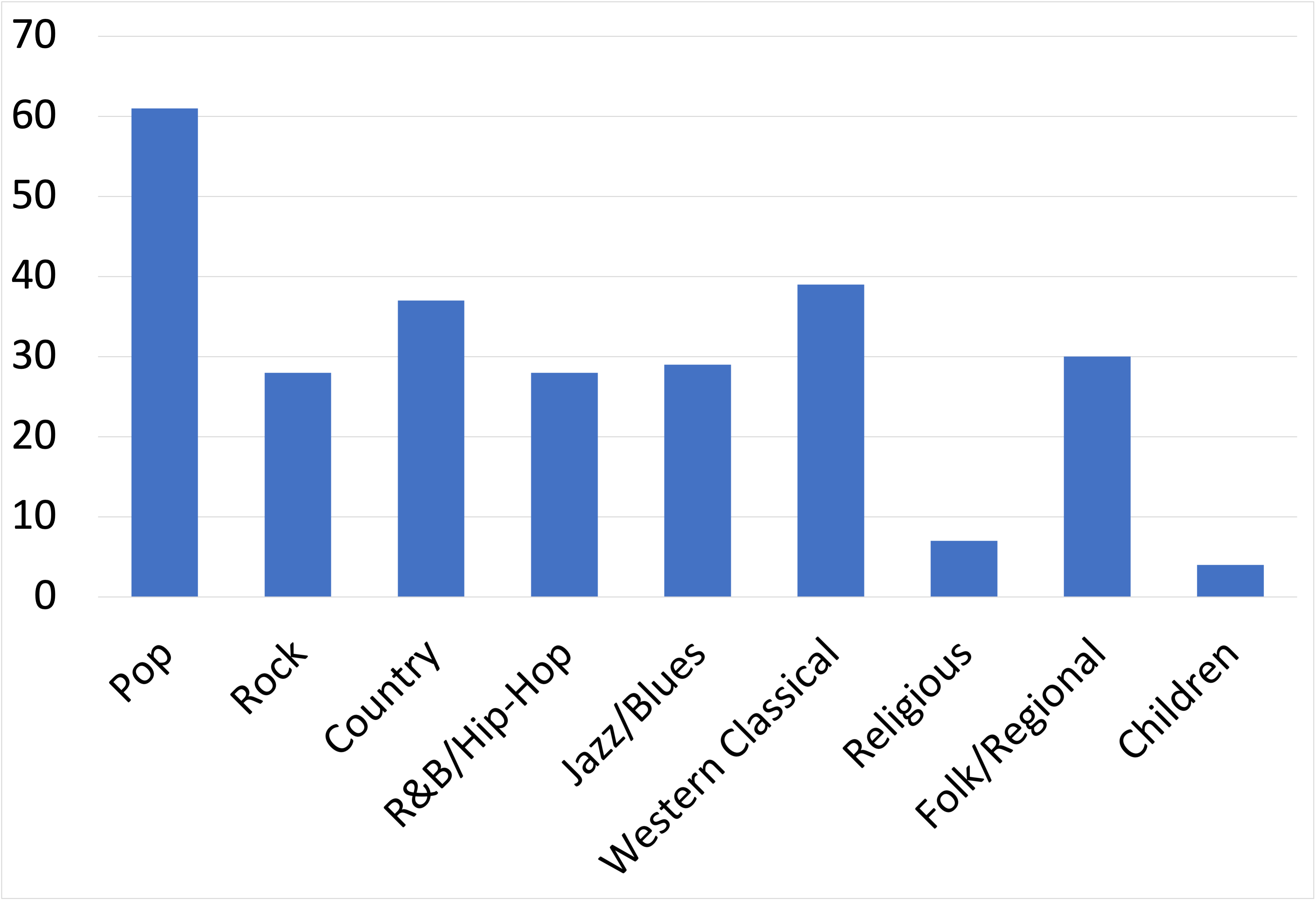}
}
\subfloat[Number of examples collected for each evaluation part]{
  \includegraphics[width=0.4\linewidth]{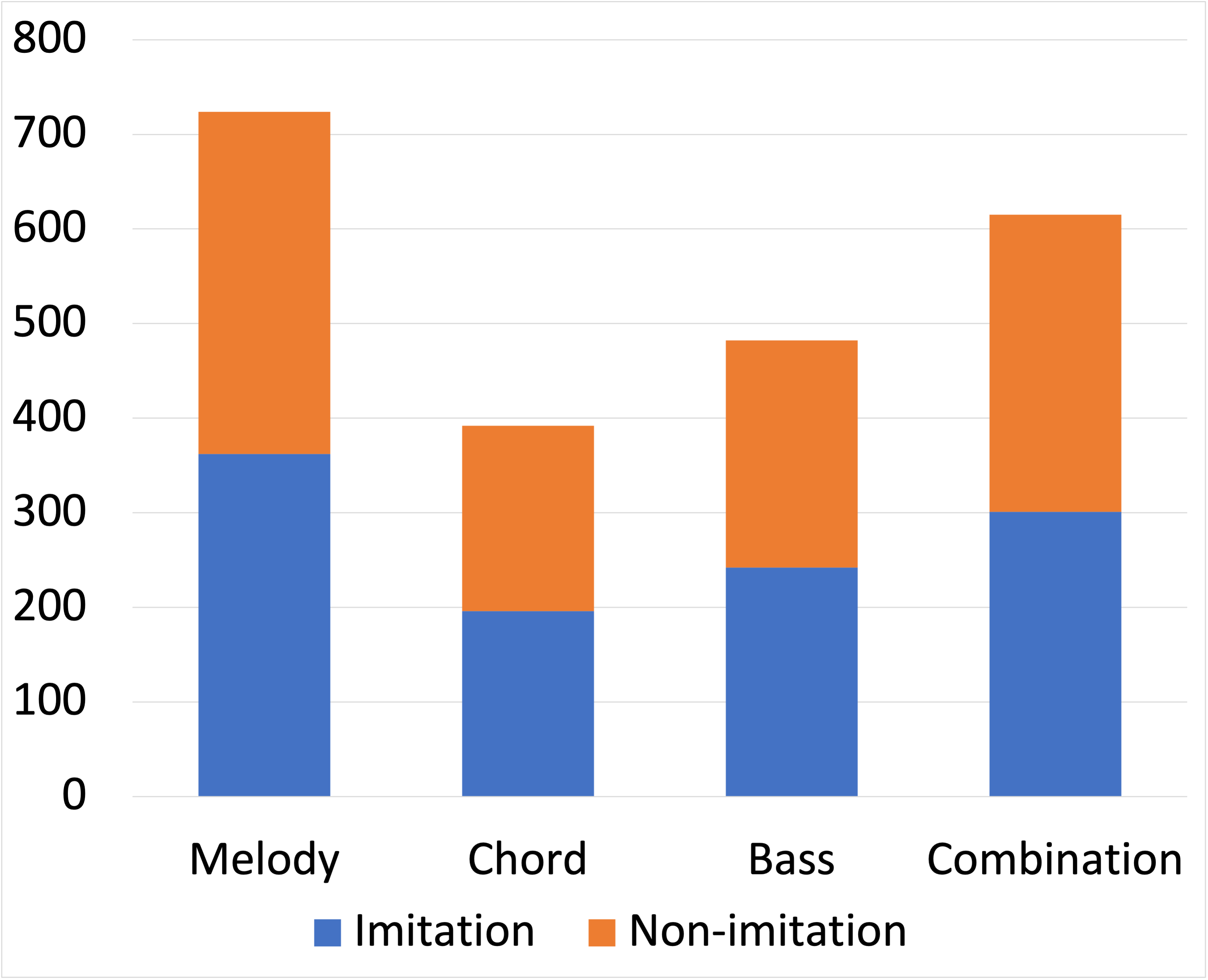}
}
\caption{Demographics Distribution}
\label{fig:demographics}
\end{figure}

We collected 154 evaluation surveys. We removed surveys that are not reliable based on validation and confidence tests. For example, some users gave different scores for the same seed songs in different parts; some users made inconsistent claims about whether they have heard the same song; some users even indicated that they have heard most of our system generated songs before. All these surveys are considered too unreliable to use. 
After removing these surveys, we are left with 2214 rating pairs from 113 listeners. The demographics information about the users are shown in Figure \ref{fig:demographics}.

\begin{figure}[!htbp]
    \centering
    \includegraphics[width=1.0\linewidth]{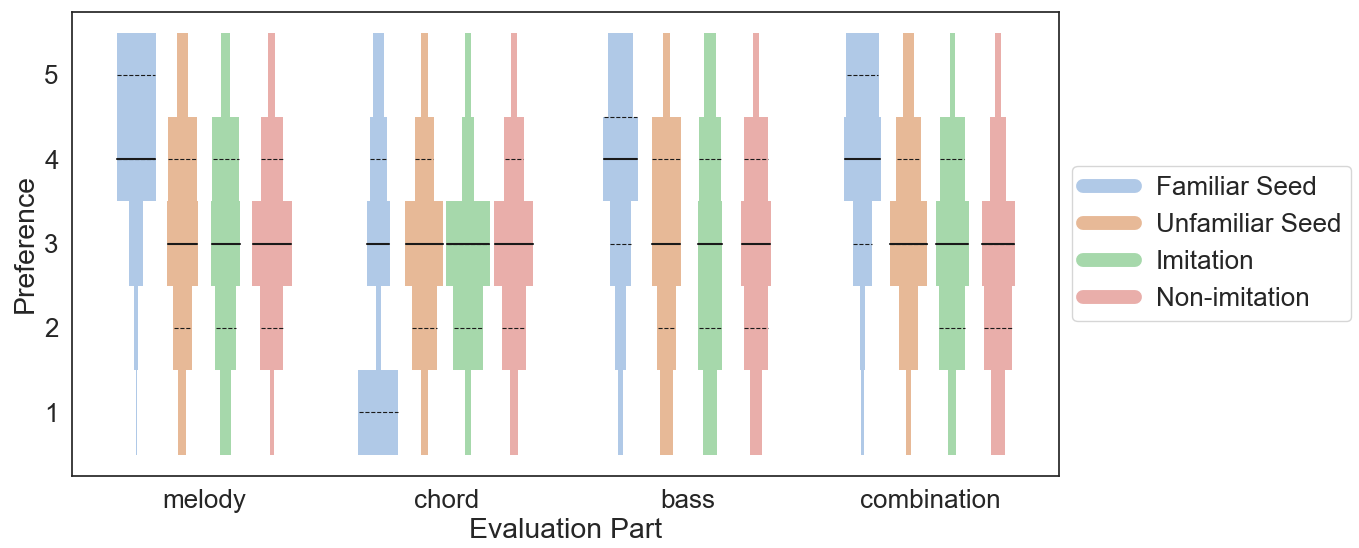}
    \caption{Preference ratings for the familiar and unfamiliar seed songs, imitations and non-imitations. Ratings of machine-generated imitations and non-imitations, when paired with familiar seed songs, were rated lower and are excluded here.}
    \label{fig:likeness}
\end{figure}

\begin{figure}[!htbp]
    \centering
    \includegraphics[width=1.0\linewidth]{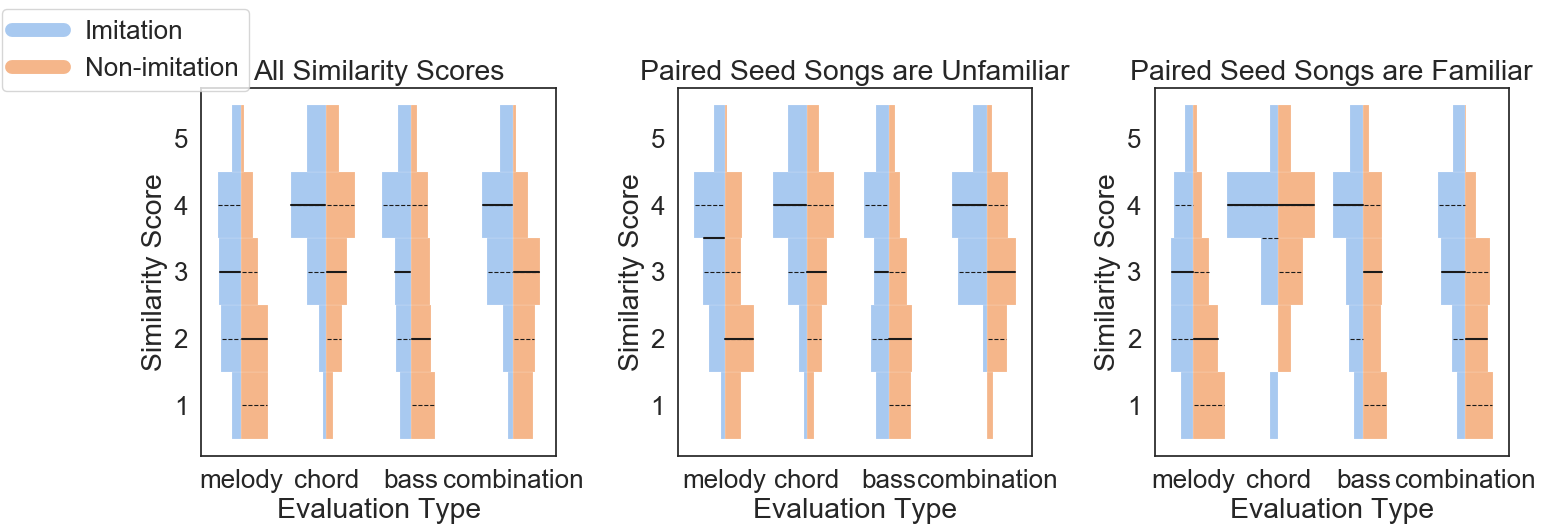}
    \caption{Similarity ratings between seed songs and generated songs, showing differences between imitations and non-imitations. The second and third plots separate the data to consider only cases where the seed songs are unfamiliar and familiar, respectively.}
    \label{fig:similarity}
\end{figure}

\begin{figure}[!htbp]
    \centering
    \includegraphics[width=1.0\textwidth]{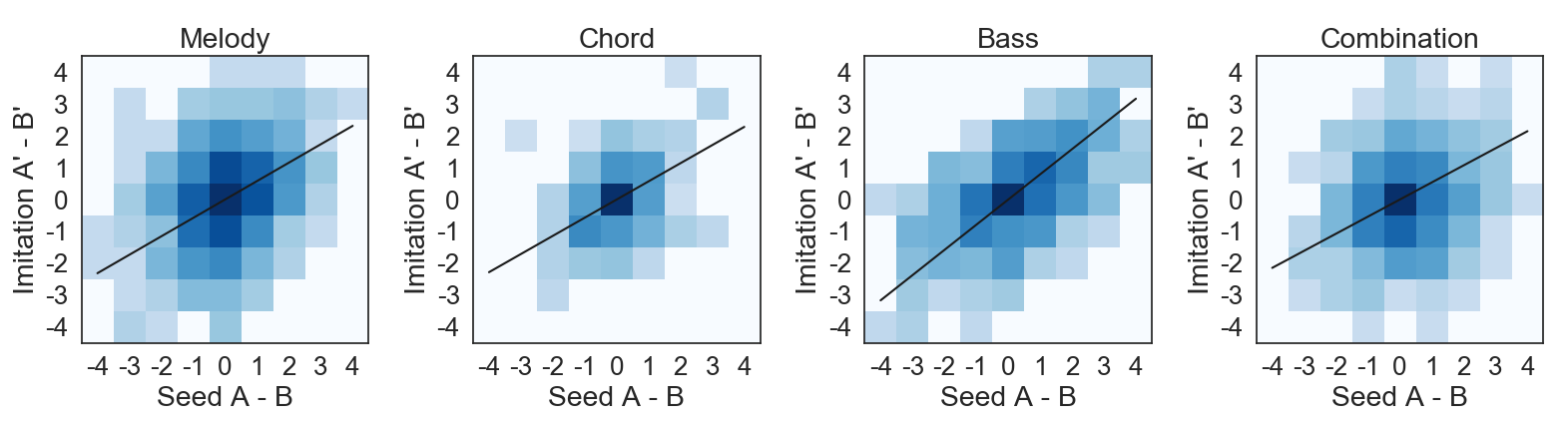}
    \caption{Correlation between preference-of-A $-$ preference-of-B and preference-of-A' $-$ preference-of-B' (all four values from the same listener), where A and B are seed songs, A' and B' are imitations. }
    \label{fig:ABcorrelation}
\end{figure}

\paragraph{Observation 1} \textit{Our system can generate good music overall.}

A problem with measuring preference is that familiar music is generally preferred over unfamiliar music \citep{pereira2011music, ward2014same}. Therefore, before the evaluations, we asked the users to indicate whether they have heard the seed songs before. In our study, based on an unpaired t-test, except for the chord progression part, listeners prefer familiar seed songs more than unfamiliar seed songs (p-value $< 0.001$), as shown in Figure \ref{fig:likeness}. Of course \textit{all} of our generated music is new and unfamiliar to listeners. Furthermore, we learned that some users consciously lowered scores for computer-generated songs when they recognized the seed song. To level the playing field, we consider only rating pairs where seed songs are not familiar. Using unpaired t-tests, we find that for all evaluation parts, if the seed song is unfamiliar, there is no significant difference between preference for the seed song and preference for either imitations or non-imitations (all p-value $> 0.05$, average p-value = $0.298$). Moreover, if the seed song is unfamiliar, $58\%$ of the imitations have the same preference ratings, and $17\%$ of the imitations have higher preference ratings, compared to their paired seed songs. Thus, the overall quality of our generated songs are usually as good, and sometimes better, than their paired seed song unless the seed song is already familiar. Non-imitations are imitations of some other (also unfamiliar) seed song, which explains why the preferences of imitations and non-imitations have no measurable differences.

\paragraph{Observation 2} \textit{The imitation techniques in our system are effective: Imitations are judged to be similar to the seed songs.}

T-tests between the similarity ratings for imitations and non-imitations show that imitations are judged significantly more similar (p-value $< 0.001$) to their paired seed songs than non-imitations, except for chord evaluation (Figure \ref{fig:similarity}). Furthermore, when users are not familiar with the seed song, the similarity of imitations for melody and combination evaluation is judged significantly higher than when the seed song is familiar (but only for melody and all features combined). Thus, it appears that familiarity affects the perception of similarity. Perhaps when listeners are more familiar with the seed song, they form strong expectations and are therefore more attentive to differences between the imitation and seed song.

\paragraph{Observation 3}\textit{The more preferable the seed song is, the more preferable the imitation will be.}

One important assumption and motivation for this research is that if we imitate the styles of the seed songs that people prefer, then they will also like the imitations. In other words, the more listeners like the seed song, the more they should like our imitation. 
To test this, consider Figure \ref{fig:seed}(a). Users rate songs in pairs of seed song and generated song, with or without imitation. We express ratings as Z-scores in order to normalize the bias and standard deviations of different raters. Let $A$ represent the preference rating for song A, and let $A'$ represent the preference rating for an imitation of A, while $B'$ represents the preference rating for a non-imitation of A. We then compute the correlation of $A$ with $B'$ as a control condition, and $A$ with $A'$ as the experimental condition. For the control condition, the correlation is significantly positive at 0.13. One possible explanation for this is that we always use the same sounds to synthesize the seed song and generated song to enable fair comparisons.  The resulting shared timbral and production qualities are likely to influence ratings. For the experimental condition, the correlation is 0.28, which is significantly higher than the control ($p < 0.001$). Therefore, we conclude that when listeners 
like seed song A more than seed song B, they like imitation A' more than non-imitation B'.

\paragraph{Feedback from Listeners}
We also received feedback and comments from the users. For example, some users said that when they did the later part the survey, they can guess there is one computer-generated song and one human-composed song in each paired example, because they are familiar with some of the human-composed seed songs. Once they figured out there is one human-composed song that they have heard before, they consciously rated the other generated song lower in the paired example. Some also told us that the chord progression is too hard for non-musical experts to evaluate, and even hard to tell the difference for professional musicians, which explains why the results for chord progression alone are not consistent with the other parts. For bass evaluation, even though we gave examples to users to explain what is a bass line, some non-musician users still said the melody and production sometimes distracted their attention from the bass when rating similarity and preference. Also, we tried to control for instrumentation and timbral preferences by using the same production rendering within each paired example, but the production might be more suitable for one genre of songs, and might lead to some bias in the rating. There are many things to consider when conducting listening tests for computer generated music, but so far there is no standard methodology. The practices we used here including confidence tests to filter out invalid evaluations, eliminating other irrelevant factors in the songs, recording whether users have listened to the full length of the songs,  listening environment control, etc., may be useful in other studies.

\section{Conclusions}
    \label{sec:conclusions}
    We have described techniques to automatically generate stylistic melody, harmony and bass lines for pop songs with a logical and hierarchical music structure. Our original motivation was to enable the creation of therapeutic music where repeated listening might call for some variety, but where therapists might not have the time, resources or permission to use existing popular music. One can imagine many other `music on demand' applications such as background music for slides and videos or interactive music for games.

We hoped that by imitating favorite songs, we could make music more personalized and enjoyable than `generic' music based on norms but without any particular style. To demonstrate this, we first needed to meet two requirements: (1) generate desirable, likeable music, (2) generate music that is similar to \textit{seed} songs. Then, we showed that (3) listeners prefer our imitations of songs they like more over imitations of songs they like less. This result shows great promise that music can be generated for a wide range of listeners and preferences using favorite songs and imitation to improve the results.

From experience, we believe that the melodic contour and rhythmic onset similarity play the most important roles in generating similar songs, at least within the probabilistic framework that we introduced. A detailed study of the influence of each of our rules on melody formation and similarity judgements, not to mention rule interactions, is beyond the scope of this study. It should also be mentioned that differences in timbre and production qualities are very important to perception, genre, and musical taste. Timbre, instrumentation and other production qualities fall outside our study of rhythm, bass, melody and harmony, so we carefully controlled for their effect in our studies. Certainly, these factors should be exploited if the goal is to enhance music preference through imitation.

In addition to our encouraging results, we are just beginning to explore the possibilities of combining different songs and genre, for example, combining the rhythm, bass, melody and harmony of as many as four different songs in order to create hybrid songs. With the addition of some editing capabilities to inject human suggestions, this could be an enjoyable creative activity, even for novices with little musical training.

We are also working to combine some of these ideas with deep learning approaches. Much of our success comes from making good choices for the next note based on probabilistic models informed by music theory and simple statistics. Perhaps more sophisticated sequence learning could make even better choices without losing the benefits we gain by incorporating knowledge of music structure and harmony in our process. We also believe that many ideas in this paper can inform future systems, serve as a baseline for comparison and evaluation, and offer insights into music perception and cognition.

\bibliographystyle{apacite}
\bibliography{BIB/paper}

\newpage
\appendix
\section{Appendix}
\subsection{User study Questionnaire}
Instructions before starting the survey:
\begin{itemize}
\item The survey will take about 50 minutes (5 parts in total). Remember to save your answers before you quit. You can resume and continue the survey anytime you want.
\item We highly recommend you do this survey on your computer and use your headphones.
\item If you meet any problem during the test, please contact us via email.
\item Remember to leave comments after the survey.
\end{itemize}

\paragraph{Part I } Music Background

\begin{itemize}
    \item Music listening and Practice (select one from the following)
    \begin{itemize}
    \item I am an expert, e.g. more than 5 years of vocal or instrumental training and practice.
    \item I studied an instrument or voice 1 to 5 years.
    \item I have less than one year study but listen at least 15 hours per week.
    \item I like music and listen 1 to 15 hours per week.
    \item I listen to music less than 1 hour per week.
    \end{itemize}
    
    \item What genre of music do you prefer? (multiple selection)
    \begin{itemize}
    \item Pop, Rock, Country,  R\&B/Hip-Hop,  Jazz/Blues,   Western Classical, Religious,\\ Folk/Regional, Children
    \end{itemize}
    
    \item What is your age? (select one from the following)
    \begin{itemize}
    \item 5 - 10, 11 - 20, 21 - 30, 31 - 40, 41 - 50, 51 - 60,  $>$ 60
    \end{itemize}
    
    \item What is your gender?  (type in)
\end{itemize}

\paragraph{Part II } Melody Evaluation (The order of Part II, III, IV is random.)

In this section, we want you to judge the similarity between \textbf{melodies} in two songs. Examples are fragments extracted from full songs. Some of them are generated by AI algorithm, and some are written by human composers. Same instructions for all examples:

Listen to \textit{Song A} and answer the following two questions, then listen to \textit{Song B} and answer the following three questions. For rating questions, try to use the \textbf{full range} from 1 to 5 stars. You must complete listening to the \textbf{full length of the song} before answering its questions. You may go back and listen to the two songs as many times as you wish.

If you are confused what is melody, listen to the example below: 

This is a song (play a song). This is its melody (play its melody).

There are 10 examples, each example is described as:

Example $n$
\begin{itemize}
    \item play \textit{Song A}
    \item Have you listened to Song A before this survey? (yes or no)
    \item How much do you like the melody of Song A? (rate 1 to 5)
    \item play \textit{Song B}
    \item Have you listened to Song B before this survey? (yes or no)
    \item How much do you like the melody of Song B? (rate 1 to 5)
    \item How similar is the melody of Song A compared to Song B? (rate 1 to 5)
\end{itemize}

\paragraph{Part III} Chord Evaluation

Similar to melody evaluation, change \textit{melody} to \textit{chords} in the instructions and questions.

\paragraph{Part IV} Bass Evaluation

Similar to melody evaluation, change \textit{melody} to \textit{bass line} in the instructions and questions.

\paragraph{Part V} Combination Evaluation

Change the first sentence of instructions to: 
In this section, we want you to judge the \textbf{overall similarity (including melody, chord and bass)} between two songs.

Change the question `How much do you like the melody/bass lines/chords of Song A/B' to `How much do you like Song A/B?'

\newpage
\pagenumbering{roman}

\end{document}